\documentclass{eds2019}

\def\etal {{\it et al.}}

\def\be{\begin{equation}}
\def\ee{\end{equation}}
\def\bea{\begin{eqnarray}}
\def\eea{\end{eqnarray}}

\def\mpm {$\pm$}

\newcommand{\Photo}{}

\begin{document}
\vspace*{4cm}
\title{RESULTS ON TOTAL AND ELASTIC CROSS SECTIONS IN PROTON-PROTON COLLISIONS AT $\sqrt{s}$ = 200 GeV OBTAINED WITH THE STAR DETECTOR AT RHIC}

\author{ For the STAR Collaboration \\ 
\vspace*{0.5cm}}
\author{ B. Pawlik }
\address{Institute of Nuclear Physics PAN, Radzikowskiego 152,\\
31-342 Cracow, Poland}
\author{ W. Guryn }
\address{Brookhaven National Laboratory, Department of Physics\\
Upton, NY 119735000, USA}

\maketitle\abstracts{We report the first results on differential, total and elastic cross sections in proton-proton collisions
at the Relativistic Heavy Ion Collider (RHIC) at $\sqrt{s}=200$ GeV.
The data were obtained  with the Roman Pot Detector subsystem of the STAR experiment.
The data used for this analysis cover the four-momentum transfer squared ~($t$) range $ 0.045 \le |t| \le 0.135$~(GeV/c)$^2$.
The Roman Pot system was placed downstream of the STAR detector. During the data taking the Roman Pots were moved to ~8$\sigma_{y}$,
the vertical distance of from the beam center. They were operated during standard data taking procedure.
The results include values of the exponential slope parameter (B), elastic cross section ($\sigma_{el}$)
and the total cross section ($\sigma_{tot}$) obtained by extrapolation of the elastic differential cross section ($d\sigma/dt$)
to the optical point at $t = 0$~(GeV/c)$^{2}$. The detector setup and analysis procedure are reviewed.
All results are compared with the world data.}

\section{The Experiment}

\begin{figure}
\begin{minipage}{0.5\textwidth}
\centerline{\includegraphics[width=1.\textwidth]{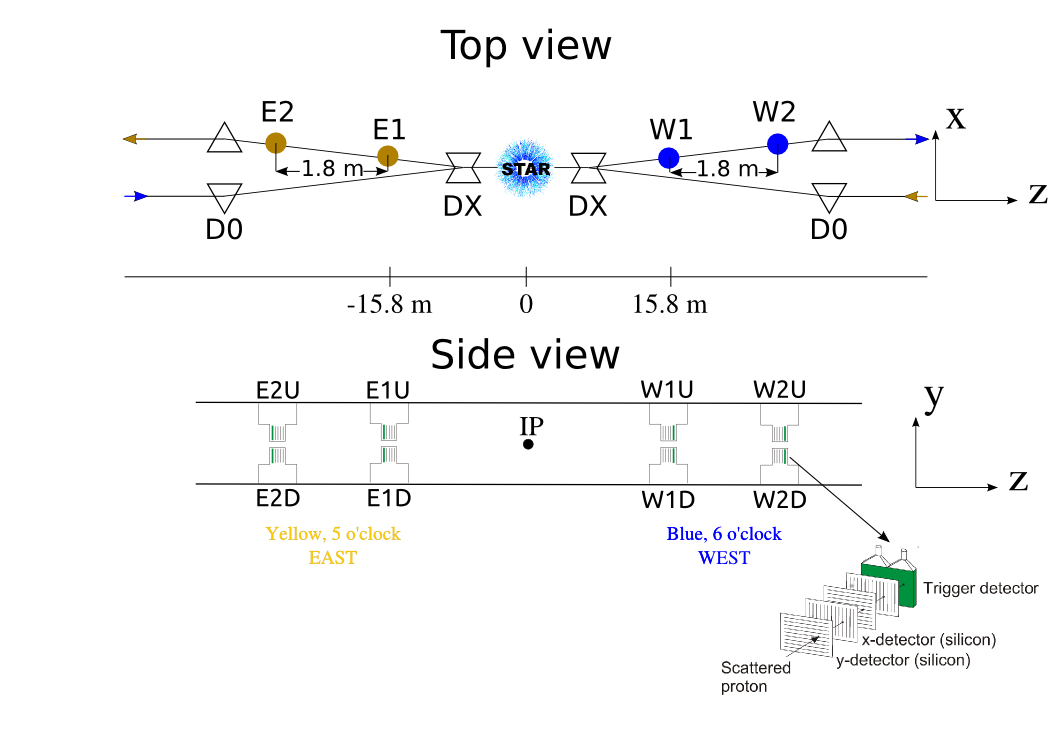}}
\end{minipage}
\hfill
\begin{minipage}{0.5\textwidth}
 \centerline{\includegraphics[width=1.\textwidth]{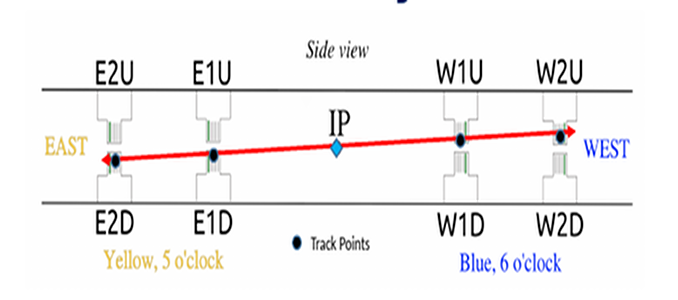}}
\end{minipage}
\caption[]{The layout of Roman Pots system at STAR (left) and example of reconstructed points configuration for elastic event detected in arm EDWU (right) .}
\label{fig:setup}
\end{figure}

Results presented here are based on data collected with pp2pp Roman Pots \cite{bib:pp2pp}
part of the STAR detector \cite{bib:STAR} at RHIC. The Roman Pots ({\bf RPs}) setup Fig.~\ref{fig:setup} consisted
of four stations with two of them ({\bf W1,W2}) placed at $\sim$15.8 and $\sim$17.6 meters downstream, (West) from interaction point ({\bf IP})
and other two ({\bf E1,E2}) symmetrically upstream, (East) from {\bf{IP}}. Each RP station consists of two Roman Pots ( one above and the other below beam line)
each equipped with package of 4 silicon strip detector (Si) planes, two planes for measuring X and the other two for Y positions of the particle track.
The scintillation counter placed behind Si planes and read by two PMTs was used to trigger on candidate events.
Candidate event had to fulfill the trigger condition, from here referred to as {\bf RP\_ET}, requiring presence of the signal in at least one Roman Pot
({\bf RP}) on each side of {\it \bf IP}.

\section{Data Set}
\begin{figure}
\begin{minipage}{0.5\textwidth}
 \centerline{\includegraphics[width=1.\textwidth]{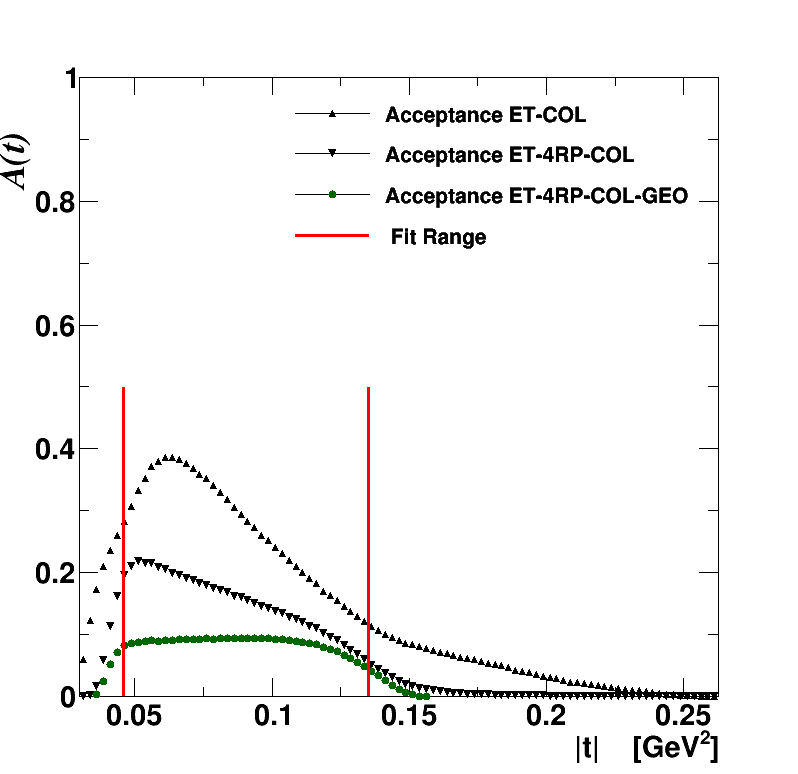}}
\end{minipage}
\hfill
\begin{minipage}{0.5\textwidth}
 \centerline{\includegraphics[width=1.\textwidth]{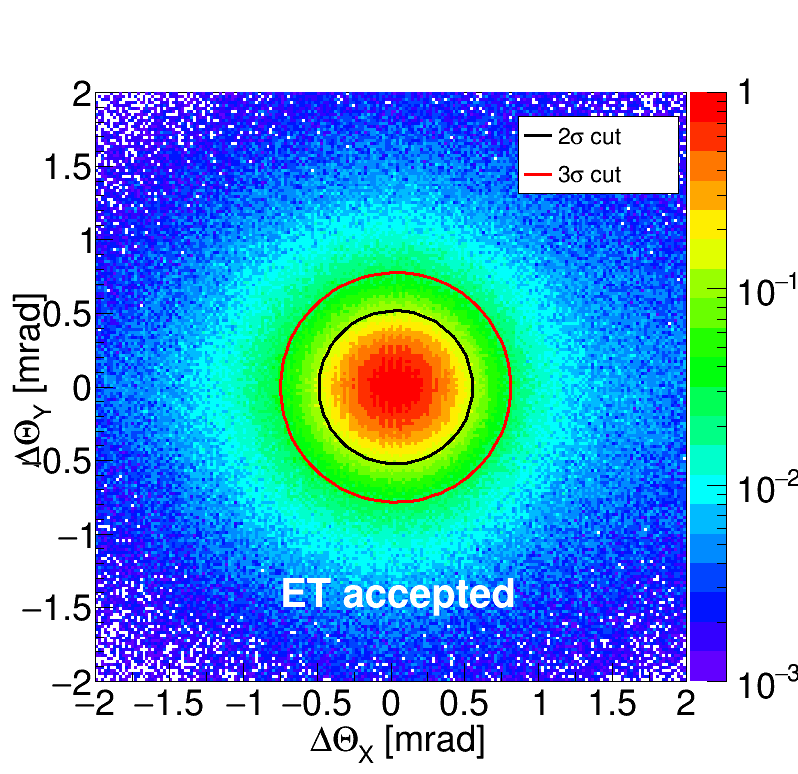}}
\end{minipage}
\caption{Acceptance as function four-momentum transfer $\mathbf t$ (left), West-East co-linearity $\Delta\theta_{Y}$ vs $\Delta\theta_{X}$ (right).}
\label{fig:accp-col}
\end{figure}

 Data were taken with nominal beam conditions $\beta^{*}=0.85m$, luminosity $\approx 45\cdotp10^{30}cm^{-2}sec^{-1}$. There were approximately 6.7 millions events
 fulfilling trigger condition {\bf\verb^RP_ET^} recorded for integrated luminosity 1.8pb$^{-1}$. The geometrical acceptance was constrained by the closest possible approach
 of the detector to the beam and, the aperture of the beam line elements (DX magnet) in front of the detector. The closest achieved distance of the first strip
 was $\sim30$ mm corresponding to minimum four-momentum transfer $|\mathbf t_{min}| \simeq $ 0.03 GeV${^2}$. The aperture of DX magnet sets the maximum achievable
 four-momentum transfer $|\mathbf{t_{max}}|\approx $0.25 GeV$^{2}$. The detector acceptance as function of four-momentum transfer $|\mathbf{t}|$ is shown
 in fig.\ref{fig:accp-col}.

\subsection{Event Reconstruction}
All events collected with trigger {\bf\verb^RP_ET^} underwent reconstruction procedure. First, in all RPs in each detector Si plane clusters - continuous set of strips with
signal above threshold - were formed. Next, clusters found in two X-planes were matched by comparing their positions $x_{1}$ and $x_{2}$
and finding the pair with minimum distance $\Delta x_{c} = |x_{1}-x_{2}| $ smaller then 200$\mu m$ (twice Si detector strip pitch).
Analogous procedure was repeated for two Y-planes. Unmatched clusters, if any, were considered as detector noise or random background and were neglected.
Pairs of clusters matched in x and y-plane defined space points $X_{RP}$ and $Y_{RP}$ coordinates of the proton track. These were used to calculate
the local angles $\theta_{x}$ and $\theta_{Y}$ in (x,z) and (y,z) planes as:
\begin{equation}
 \theta_{X} = \frac{X_{RP1}-X_{RP2}}{Z_{RP1}-Z_{RP2}} ~~~ and ~~~  \theta_{Y} = \frac{Y_{RP1}-Y_{RP2}}{Z_{RP1}-Z_{RP2}}
 \label{eq:angles}
\end{equation}
\noindent where subscripts RPs(RPs) denote RP stations 1(2) at same side of $IP$ and  $Z_{RP1}$($Z_{RP2}$) are z-positions of the stations.
For small scattering angles in this experiment, to a good approximation the four-momentum transfer $t$ was calculated with formula:
\begin{equation}
 t = -p^{2}\cdotp \theta^{2} = -p^{2}\cdotp( \theta^{2}_{X} + \theta^{2}_{Y})
 \label{eq:tequ}
\end{equation}
\noindent where $p$ is proton momentum, $\theta = \sqrt{\theta^{2}_{X}+\theta^{2}_{Y}}$ scattering angle
and $\theta_{X}, \theta_{Y}$ calculated as in eq.\ref{eq:angles}. The RPs system was positioned and aligned with respect to nominal beam
trajectory, hence the angles $\theta_{X}$ and $\theta_{Y}$ provide direct measurement of the projections of scattering angle $\theta$ on (x,z) and (y,z) plane,
respectively.

\subsection{Elastic Scattering Event Selection}

The hardware trigger requiring signal in at least one RP on each side of IP was very inclusive.
The clean pattern indicating elastic scattering ( see right sub-figure in fig.\ref{fig:setup} ) is presence of two back to back protons in the event.
This requires signal only in top RP1 and/or RP2 at one side of $IP$ and only in bottom RP1 and/or RP2 on the other side.
Calculation of the track direction angles (eq.\ref{eq:angles}) requires points in two stations on the each side of IP.

\begin{figure}
\begin{minipage}{0.5\textwidth}
\centerline{\includegraphics[width=1.\textwidth]{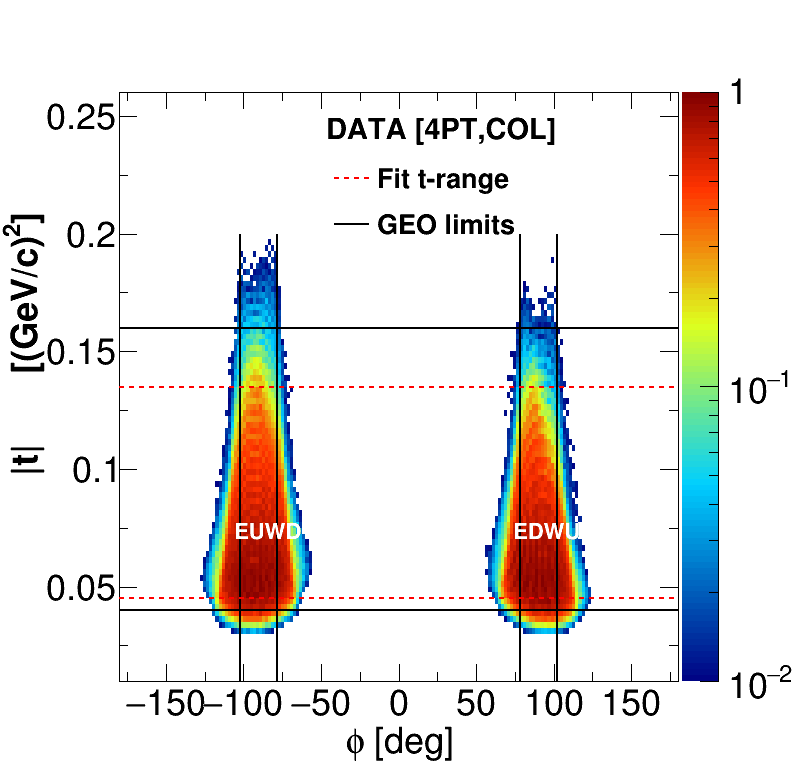}}
\end{minipage}
\hfill
\begin{minipage}{0.5\textwidth}
 \centerline{\includegraphics[width=1.\textwidth]{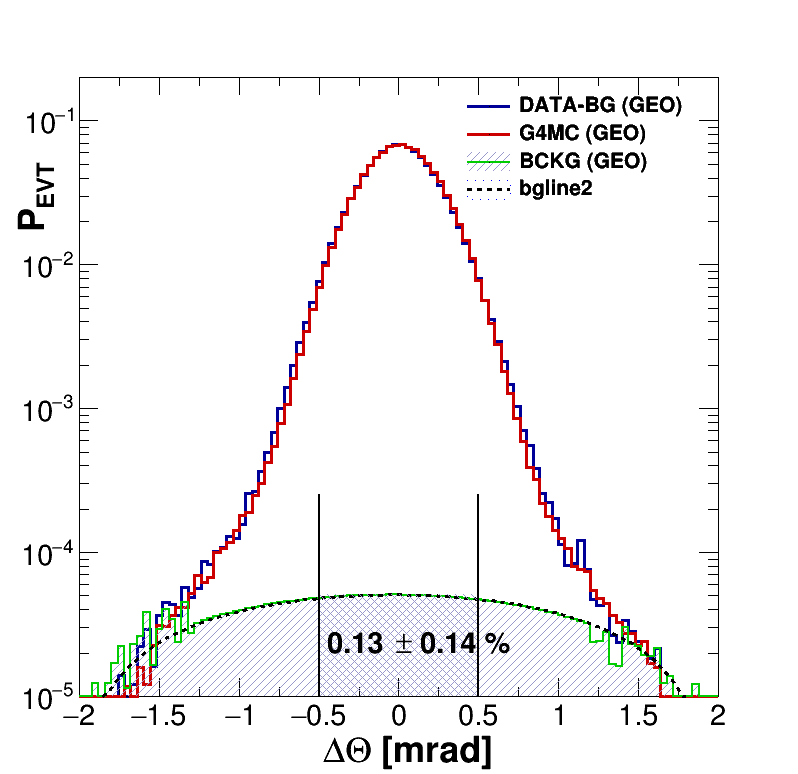}}
\end{minipage}
\caption{Four momentum transfer $|t|$ vs azimuth angle $\phi$ for accepted ET co-linear events with four reconstructed points ({\bf 4PT}) (left),
 and (right) background contribution estimate based on comparison of West-East co-linearity $\Delta\theta$ for DATA and Monte-Carlo samples of {\bf 4PT} events
 within {\bf GEO} limits (\ref{eq:geocut}). }
\label{fig:geo-accp-bckg}
\end{figure}

The data sample used to obtain this results consist only of events with four reconstructed points, four points ({\bf4PT}) events, and fulfilling West-East co-linearity
condition:
\begin{equation}
\Delta\theta =\sqrt{ (\theta^{West}_{X}-\theta^{East}_{X})^{2} + (\theta^{West}_{Y}-\theta^{East}_{Y})^{2} } < 2\cdotp \sigma_{\theta}
\label{eq:colinear}
\end{equation}
\noindent with $\sigma_{\theta}=255~\mu$rad was dominated by the beam angular divergence ($\sim 180~\mu$rad for each beam). The kinematic range of four-momentum
transfer $\mathbf{t}$ versus azimuth angle $\phi$ for this sample ({\bf4PT-COL}) is shown in fig.\ref{fig:geo-accp-bckg}.

For the 4PT-COL events, scattering angles at the IP $\theta_{X}^{*}, \theta_{Y}^{*}$ were obtained from the linear fit using four $X_{RP}$ and $Y_{RP}$ points.
The four-momentum transfer t for those events was then calculated using Eq.\ref{eq:tequ}, where local angles $\theta_{X}, \theta_{Y}$ were replaced respectively
with $\theta^{*}_{X}, \theta^{*}_{Y}$.

Additionally geometrical cut was imposed to reduce background by staying  away from acceptance boundaries and maintain relatively flat, slow varying
acceptance corrections (see data labeled as {\bf ET-4RP-COL-GEO} in fig.\ref{fig:accp-col}).
It was required that the scattered proton angle $\theta$ and azimuth angle $\phi$ obey following limits:
\begin{equation}
 79.5 \texttt{[deg]} < | \phi | < 101.5 \texttt{[deg]}  ~~~  2.0 \texttt{[mrad]} < \theta < 4.0 \texttt{[mrad]}
\label{eq:geocut}
\end{equation}

\subsection{Monte Carlo Corrections}
The beam line elements and all RP detectors were implemented in detail in Geant4$^{[}$\cite{bib:Geant4}$^{]}$ based Monte Carlo
application. The events were generated according to standard formula for the elastic scattering differential cross section with
the slope $B = 14.0$ GeV$^{-2}$, the parameter $\rho$=0.128 and West-Yennie \cite{bib:y-w} Coulomb phase. The beam angular divergence and
the interaction point $\mathbf{IP}$ position uncertainty were included in the generator.

The experimental differential distributions $dN/dt$ was corrected using ``bin by bin'' method with the formula :
\begin{equation}
 \bigg ( \frac{dN}{dt} \bigg )^{DATA}_{corrected} = \bigg ( \frac{dN}{dt} \bigg )^{DATA}_{reconstructed} \times \frac{(dN/dt)^{MC}_{generated} }{(dN/dt)^{MC}_{reconstructed}}
 \label{eq:CorrectionForm}
\end{equation}
\noindent where $(dN/dt)^{MC}_{generated}$ and $(dN/dt)^{MC}_{reconstructed}$ are true MC distribution and  reconstructed based on MC event
sample which passed the same reconstruction procedure and selection criteria as those applied for experimental data. The corrections obtained this way
account for limited geometrical acceptance, effects of the scattering angle reconstruction resolution ( $\mathbf{t}$ smearing ) and impact of the secondary scattering
of the final state proton off the material on the way from $\mathbf{IP}$ to detector Si planes.

\section{Results}
\begin{figure}
 \centering
 \includegraphics[width=0.5\textwidth]{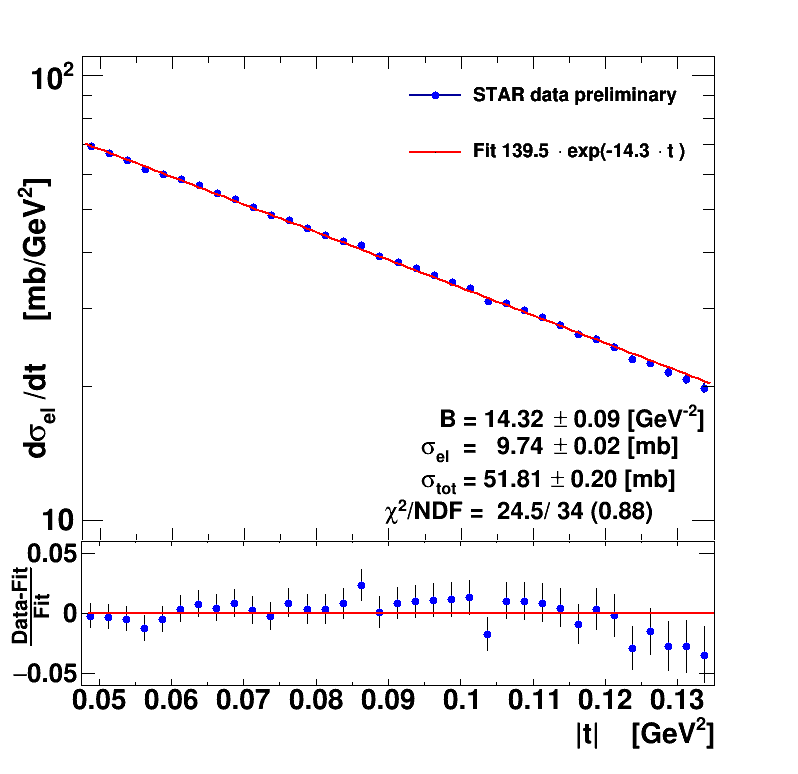}
 \caption{ Top panel: $pp$ elastic differential cross-section $d\sigma/dt$ fitted with exponential $A\cdot exp(-Bt)$; Bottom panel: Residuals (Data - Fit)/Data.}
\label{fig:bfit}
\end{figure}

The corrected differential cross section ($ d\sigma/dt$ ) was fitted with standard formula \cite{bib:Kopel,bib:Barone,bib:Donna} :
\begin{equation}
 \frac{d \sigma_{el}}{dt} = \frac{1 + \rho^{2}}{16 \pi (\hbar c)^{2}} \cdot \sigma^{2}_{tot} \cdot e^{-B|t|}
 \label{eq:sig}
\end{equation}
with $\rho$=0.128 from COMPETE \cite{bib:COMPETE} model. The Coulomb and interference terms were neglected as their contribution
in the fit range 0.045$< -t <$ 0.135 GeV$^{2}$ is negligibly small within this experiment's precision. The data and fit results are shown in Fig.\ref{fig:bfit}.

 The total cross section  $\sigma_{tot}$ was calculated using the optical theorem as :
 \begin{equation}
  \sigma_{tot}^{2} = \frac{16 \pi (\hbar c)^{2}}{1 + \rho^{2}} \cdot \frac{d\sigma_{el}}{dt}\texttt{\huge|}_{t=0}
 \end{equation}
 and the total elastic cross section $\sigma_{el}$ was obtained by integrating fitted formula (\ref{eq:sig}) over whole $t$ range, the elastic cross section integrated
 within the t-acceptance of this measurement ($\sigma^{det}_{el}$) is also quoted.
 The inelastic cross section  is simply result of subtraction $\sigma_{inel}$=$\sigma_{tot} - \sigma_{el}$.
 All results with their statistical and systematic uncertainties are shown in table~\ref{tab:FinSysErr}.

\begin{table*}
 \caption{ Results summary. }
 \label{tab:FinSysErr}
\begin{minipage}{0.5\textwidth}
 \centering
 \begin{tabular}{c|c|r|c|c|c|c|c}
 \hline \hline
 \multicolumn{3}{ c|}{Quantity}          & Statistical & \multicolumn{4}{|c}{ Systematic uncertainties}\\
 \cline{1-3} \cline{5-8}
  name        & units            & Value & uncertainty & beam-tilt& luminosity & $\rho$ & full \\ [1ex]
 \hline \\ [-2.5ex]
 $d\sigma_{el} /dt \vert _{t=0}$ & [mb/GeV$^{2}$]    & 139.53 & \mpm 1.06   & $^{+1.07}_{-0.83}$ & $^{+10.50}_{-10.07}$  &   n/a            & $^{+10.55}_{-10.10}$ \\ [1ex]
 \textbf{B}  & [GeV$^{-2}$]      &  14.32 & \mpm 0.09   & $^{+0.18}_{-0.32}$ &         n/a           &   n/a            & $^{+0.18}_{-0.32}$  \\ [1ex]
  $\sigma_{el}$     & [mb]       &   9.74 & \mpm 0.02   & $^{+0.06}_{-0.04}$ & $^{+0.74}_{-0.59}$    &   n/a            & $^{+0.74}_{-0.59}$  \\ [1ex]
  $\sigma^{det}_{el}$ & [mb]     &   3.63 & \mpm 0.01   & $^{+0.02}_{-0.01}$ & $^{+0.28}_{-0.23}$    &   n/a            & $^{+0.28}_{-0.23}$  \\ [1ex]
  $\sigma_{tot}$    & [mb]       &  51.81 & \mpm 0.20   & $^{+0.19}_{-0.61}$ & $^{+1.91}_{-1.90}$    & $^{+0.20}_{-0.40}$ & $^{+1.93}_{-2.04}$  \\ [1ex]
  $\sigma_{inel}$   & [mb]       &  42.07 & \mpm 0.20   & $^{+0.20}_{-0.61}$ & $^{+2.05}_{-1.99}$    & $^{+0.20}_{-0.40}$ & $^{+2.07}_{-2.12}$  \\
 \hline
 \end{tabular}
 \end{minipage}

 \end{table*}

\section{Summary}
The elastic differential cross section in {\it pp} scattering was measured with Roman Pots system of the STAR experiment at RHIC
in $t$ range 0.045$< -t <$ 0.135 GeV$^{2}$ at $\sqrt{s}$=200 GeV. In this $t$ range the cross section is well described by exponential
$exp(-B\cdot t)$ with the slope B = 14.32$\pm$0.09($^{+0.18}_{-0.32}$) GeV$^{-2}$, in brackets full systematic errors are given.
The elastic cross section integrated within detector acceptance $\sigma^{det}_{el}$ = 3.63$\pm$0.01($^{+0.28}_{-0.23}$) mb, extrapolation of
this measured cross section over undetected (~60\%) $t$ region results in value of the total elastic cross section $\sigma_{el}$=9.74$\pm$0.02($^{+0.74}_{-0.59}$)mb.
Using optical theorem we found the value of total $pp$ scattering cross section $\sigma_{tot}$=51.81$\pm$0.20($^{+1.93}_{-2.04}$).

\begin{figure}[h]
\begin{minipage}{0.5\textwidth}
\centerline{\includegraphics[width=\textwidth]{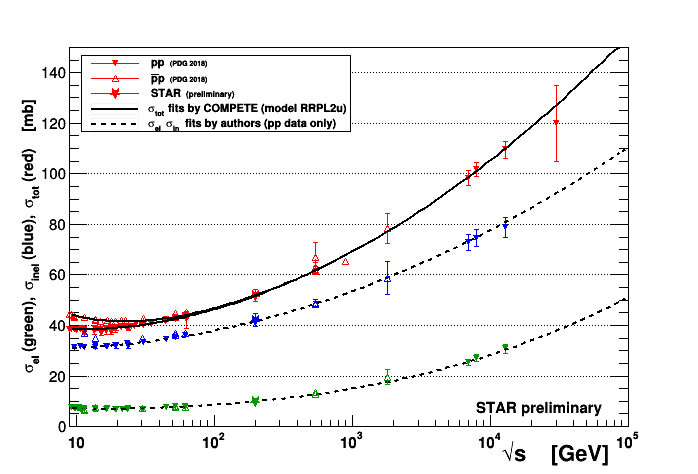}}
\end{minipage}
\hfill
\begin{minipage}{0.5\textwidth}
\centerline{\includegraphics[width=\textwidth]{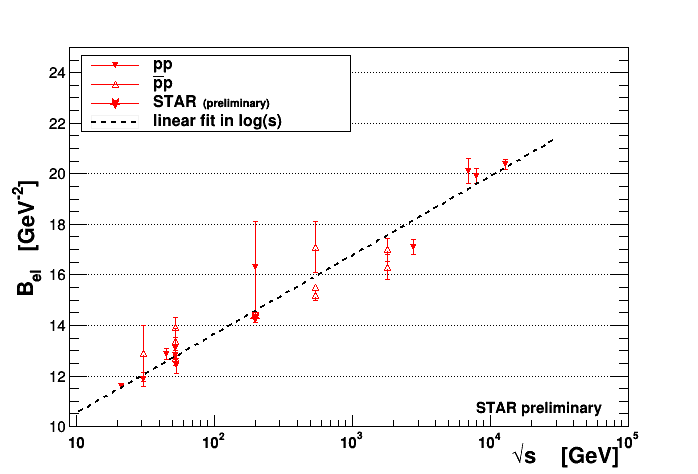}}
\end{minipage}
\caption[]{Comparison of the STAR result on $\sigma_{tot}, \sigma_{el}$ and $\sigma_{inel}$ (left) and B-slope (right) with the world data
on cross sections \cite{bib:PDG} and B-slopes \cite{bib:barbie,bib:ambrosio,bib:amos,bib:bueltmann,bib:totem1,bib:totem2,bib:atlas1,bib:atlas2},
COMPETE prediction \cite{bib:COMPETE} for $\sigma_{tot}$ and $\sigma_{inel}$ are displayed. }
\label{fig:WorldComp}
\end{figure}

The results obtained with STAR are compared with the world data in Fig.\ref{fig:WorldComp}.
We found they compare well and follow COMPETE prediction of dependence of cross section on $\sqrt{s}$.

\section*{Acknowledgments}
This work was partly supported by the National Science Center of Poland
under grant number UMO-2015/18/M/ST2/00162.

\end{document}